# Quantum Superposition, Collapse of Wave Function, Quantum Measurement and Nonadiabatic Dressed States

I. G. Koprinkov

*Department of Applied Physics, Technical University of Sofia, 1000 Sofia, Bulgaria*

*igk@tu-sofia.bg*

**Abstract.** Quantum superposition, collapse of wave function and quantum measurement problem are reexamined based on nonadiabatic dressed states and experimental observations on the quantum transitions. The physical mechanisms behind these processes are revealed.

## 1. INTRODUCTION

The quantum mechanics is highly successful theory that may describe wide range of physical phenomena, from microscopic to macroscopic ones, including even some cosmological problems. The areas in which the quantum mechanics really shines are those concerning atoms, molecules, condensed matter. Based on it, we have a detailed knowledge about the structure of the matter, the quantum states and the respective energy structure, the physical phenomena that may develop within these states and energy structures, the right set of parameters in which they can be quantified, etc. Apart from these, some fundamental physical problems still become open as collapse of wave function, quantum measurement, quantum entaglement, etc.

*The collapse of the wave function* (or the state vector), takes place when the wave function, being, in general, in a linear superposition of eigenstates of the quantum system, underdoes a *sudden*, *discontinuous*, *irreversible*, *non-unitary* evolution and thus shrinks onto one of the eigenstates of the quantum system due to the interaction with the environment, in particular, with the measuring apparatus in the process of observation/measurement: $|\psi\rangle = \sum_i c_i |\phi_i\rangle \rightarrow |\varphi_k\rangle$. It is also known as the von Neumann's "process 1" [1], As a result of the measurement, the eigenvalue of the observable that correspond to that eigenstate will be measured. The collapse of the wave function cannot be described by the Schrödinger equation. There is a second type of evolution of the state of the quantum system, the von Neumann's "process 2", which is *continuous*, *reversible*, *causal*, *unitary* evolution ruled by the Schrödinger equation. The collapse of the wave function is the essence of *the quantum measurement*. The evolution of the wave function in the "process 2" leads, in general, to a linear superposition of eigenstates with many possible measuring values of given observable, whereas, the "process 1", *i.e.*, the measurement, always gives a definite value of the physical observable. The inability to observe (till now) such an instantaneous collapse and the lack of mechanism that explains it are the reasons to arise the so called *quantum measurement problem*. An important step toward the understanding of the evolution of wave function and, particularly, to explain the collapse of wave function is the theory of decoherence [2, 3]. However, as has been recently recognized, the decoherence can explain transition from quantum to classical but cannot explain the collapse of wave function onto a single eigenstate [4].

The aim of the present work is to propose a *physical mechanism* for the quantum superposition, based on the dynamics of a quantum system interacting with an electromagnetic field and to explain on that ground the collapse of wave function and the quantum measurement problem. The problem is treated in terms of nonadiabatics dressed states [5] which involve the interaction of a quantum system with a generally nonadiabatic electromagnetic field and the environment considered as nonadiabatic factors that are responsible for quantum transitions.

## 2. NONADIABATIC DRESSED STATES

The following types of states are involved in the present considerations: bare states (BS), adiabatic (dressed) states (ADS) [6-8] and nonadiabatic dressed states (NADS) [5]. The BS are the states of completely isolated, *i.e.*, *closed quantum system*. The BS are the most widely used quantum states, *e.g.*, the wave functions of the hydrogen atom, the harmonic oscillator, etc. The ADS are the states of a quantum system in the presence of classical adiabatic (slowly varying) electromagnetic field. These states are originally called adiabatic states, but, as they are considered equivalent [6] (to certain extent) to their full quantum analog [9], they will be called here ADS. Finally, the NADS [5] are states of *open quantum system* in the presence of nonadiabatic factors from the electromagnetic field and environment, including zero point vacuum fluctuations. The NADS are *nonadiabatic* and *nonperturbative* simultaneously. NADS become generalization of the ADS and the BS. If we neglect all nonadiabatic factors from the field and the environment, the NADS are transformed into ADS. If we also totally neglect the field, the NADS and the ADS are transformed into BS. The following notations for the ground and excited BS, ADS, NADS will be used here: $|g\rangle$ and $|e\rangle$ ; $|G\rangle$ and $|E\rangle$ ; $|\tilde{G}\rangle$ and $|\tilde{E}\rangle$, respectively.

The physical grounds and some mathematical details of the ANDS will be presented shortly. We consider an open two-level quantum system without internal degrees of freedom (atomic type system) having allowed electric dipole transition between the initial BS, Fig.1. The quantum system is subject to a "regular" interaction with an external electromagnetic field and a stochastic interaction with the environment (collisions with other particles, zero-point vacuum fluctuations, etc.). The total Hamiltonian $\hat{H}$ of the quantum system consists of Hamiltonian of quantum system itself $\hat{H}_0$, Hamiltonian of interaction of quantum system with external electromagnetic field $\hat{H}'$ and Hamiltonian of interaction with environment $\hat{H}_D$, described by damping rates $\gamma_j$:

$$\hat{H} = \hat{H}_0 + \hat{H}' + \hat{H}_D = \sum_{j=1}^{2}\hbar\omega_j |j\rangle\langle j| - \mu E(t)(|1\rangle\langle 2| + h.c.) - i\hbar\sum_{j=1}^{2}\frac{\gamma_j}{2}|j\rangle\langle j| \quad , \tag{1}$$

where the states $|1\rangle \equiv |g\rangle$ and $|2\rangle \equiv |e\rangle$ are the ground and excited BSs of the non-perturbed quantum system, respectively. In the notations $|g\rangle$ and $|e\rangle$, all characteristics of the states (quantum numbers, symmetries, etc.), are included and will be called shortly $|g\rangle$-type and $|e\rangle$-type states. The electromagnetic (optical) field is near resonant and presented in carrier-envelope form

$$E(t) = (1/2)E_o(t)[\exp(i\Phi_F(t)) + c.c.] \quad , \tag{2}$$

where $E_o(t)$ and $\Phi_F(t) = \omega t + \varphi(t)$ are the amplitude/envelope and the total phase of the field, respectively, considered as arbitrary functions of time, subject only to a generalized adiabatic condition [5] (see below), $\omega$ is the carrier frequency, and $\varphi(t)$ is the phase of the field. The above representation holds for very broad range of fields ranging from the simplest monochromatic continuous waves to ultrashort pulses of envelope duration as short as the carrier wave period, which is in the order of few femtoseconds in the optical range. It allows involving the ultrafast phenomena, which are expected to be strongly influenced by the nonadiabatic effects.

The NADS are derived from an analytic solution of the time dependent Schrödinger equation with Hamiltonian, Eq. (1), following the procedure described in [5]:

$$\hat{H}|\Psi(\vec{r},t)\rangle = i\hbar\partial_t|\Psi(\vec{r},t)\rangle \tag{3}$$

Each NADS, ground $|\tilde{G}\rangle$ and excited $|\tilde{E}\rangle$, consists of real (index "r") and virtual (index "v") components:

$$|\tilde{G}\rangle = COS\left(\frac{\theta}{2}\right)|\tilde{G}_r\rangle + SIN\left(\frac{\theta}{2}\right)|\tilde{G}_v\rangle \tag{4}$$
$$|\tilde{E}\rangle = COS\left(\frac{\theta}{2}\right)|\tilde{E}_r\rangle - SIN\left(\frac{\theta}{2}\right)|\tilde{E}_v\rangle$$

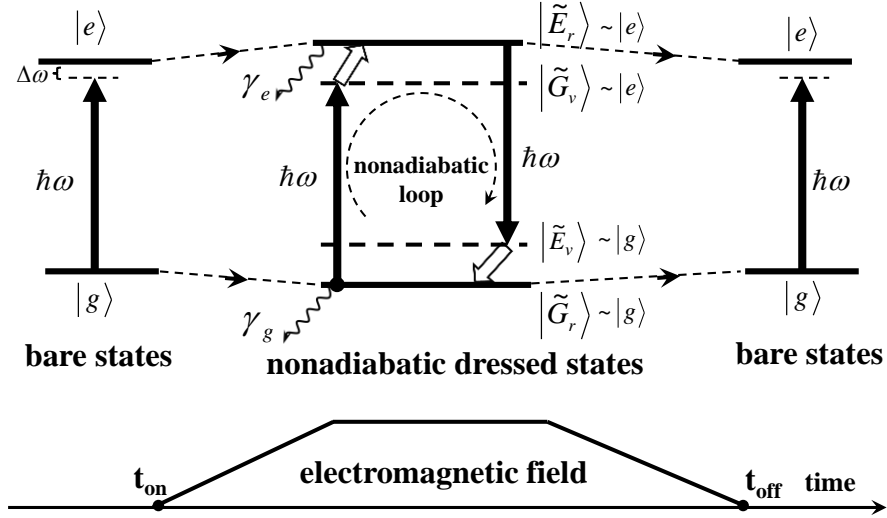

**FIGURE 1.** Evolution of bare states toward nonadiabatic dressed states and back to bare states with switching on/off $t_{on}/t_{off}$ of electromagnetic field (bottom) and damping. The real and virtual components of ground and excited nonadiabatic dressed states and their "symmetry" properties are shown. Solid arrows show radiative transitions, hollow arrows show nonadiabatic transitions.

The real and the virtual components of the NADS (at, *e.g.*, ground state initial conditions) can be expressed as

$$\begin{aligned}
|\tilde{G}_r\rangle &= |g\rangle \exp\left[-i\int_0^t \tilde{\omega}'_G\, dt'\right] \\
|\tilde{G}_v\rangle &= |e\rangle \exp\left[-i\int_0^t (\tilde{\omega}'_G+\omega)dt' - i\varphi(t)\right] \\
|\tilde{E}_r\rangle &= |e\rangle \exp\left[-i\int_0^t \tilde{\omega}'_E\, dt' - i\varphi(t)\right] \\
|\tilde{E}_v\rangle &= |g\rangle \exp\left[-i\int_0^t (\tilde{\omega}'_E-\omega)dt'\right]
\end{aligned} \qquad (5)$$

The quantities $\cos(\theta/2)=\sqrt{\tilde{\Lambda}'_1/\tilde{\Omega}'}$ and $\sin(\theta/2)=\mathrm{sgn}(\Delta\omega)\sqrt{-\tilde{\Lambda}'_2/\tilde{\Omega}'}$ are complex functions that determine the partial contribution of real and virtual components in given NADS, $\tilde{\Omega}'=\mathrm{sgn}(\Delta\omega)[\Omega^2 + \Delta\tilde{\omega}'^2 - i2\partial_t\Delta\tilde{\omega}']^{1/2}$ is nonadiabatic Rabi frequency, $\Delta\tilde{\omega}' = \Delta\omega - i(\gamma_g+\gamma_e)/2 - (\partial_t\varphi - i\,\Omega^{-1}\partial_t\Omega)$ is nonadiabatic frequency detuning, $\Lambda_1 = (\Delta\tilde{\omega}'+\tilde{\Omega}')/2$, $\Lambda_2 = (\Delta\tilde{\omega}'-\tilde{\Omega}')/2$ and $\tilde{\Lambda}'_j = \Lambda_j - i(2\tilde{\Omega}')^{-1}\partial_t\tilde{\Omega}'$ are nonadiabatic quantities associated with the Stark shift, $\tilde{\omega}'_G = \omega_g + \Lambda_2$ and $\tilde{\omega}'_E = \omega_e - \Lambda_2 - i(\gamma_g+\gamma_e)/2 - (\partial_t\varphi - i\,\Omega^{-1}\partial_t\Omega)$ are nonadiabatic frequencies ("energies") of the ground and the excited NADS, respectively, and $\Omega(t)=\mu E_o(t)/\hbar$ and $\Delta\omega = \omega_e - \omega_g - \omega$ are the usual Rabi frequency and frequency detuning, respectively. The time derivatives of the field amplitude/Rabi frequency and phase and the damping rates quantify the nonadiabatic factors, which spread out in all parameters of the quantum states.

Generalized adiabatic condition has been introduced in the derivation of the NADS [5]:

$$\left|\partial_t^n(\partial_t\varphi - i\Omega^{-1}\partial_t\Omega)\right| \ll \left|\Delta\omega - i\gamma/2\right|^{n+1-k}|\Omega|^k, \qquad (6)$$

where $n = 0, 1, 2,\ldots$; $k = 0, 1, 2,\ldots, n+1$. It represents an adiabatic condition of infinite order, which unifies and generalizes the ordinary adiabatic condition, $\left|E_o^{-1}\partial_t E_o(\Delta\omega - i\gamma)^{-1}\right| \ll 1$ [8] at $n = 0, k = 0$, and Born-Fock adiabatic condition, $\left|\partial_t\Omega^{-1}\right| \ll 1$ at $n = 0, k = 1$. The nonadiabatic factors are retained (up to given order) in the NADS, in contrast to the ADS and the BS, where they are completely eliminated.

# 3. QUANTUM SUPERPOSITION, COLLAPSE OF WAVE FUNCTION, QUANTUM MEASUREMENT PROBLEM IN NONADIABATIC DRESSED STATES PICTURE

In order to understand the problem with quantum superposition, collapse of wave function, quantum measurement, quantum jumps, etc., we will first analyze the physical properties of the real and virtual components of the NADS, and, on that ground, to realize how the states participate in the quantum superposition.

## 3.1 Physical Properties of Nonadiabatic Dressed States

The NADS have same structure as the ADS, but include in addition nonadiabatic factors, which have crucial role for quantum transitions. Each NADS, ground and excited, consist of real and virtual components. The real component originates from the respective BS, Fig.1, subject to perturbation (Stark shift and broadening) from the field and the environment. The virtual component originates from the interaction of the quantum system with the field. The energy of the virtual state is defined by the energy of the real state, from which the respective virtual state originate, plus energy of one photon from the field, temporary associated to the quantum system. The virtual component of given NADS has "opposite" characteristics to these of real component of the same NADS due to electric dipole interaction of the quantum system with the field. Thus, while the real ground state $|\tilde{G}_r\rangle$ is $|g\rangle$-type state, the virtual ground state $|\tilde{G}_v\rangle$ is $|e\rangle$-type state, respectively, while the real excited state $|\tilde{E}_r\rangle$ is $|e\rangle$-type state, the virtual excited state $|\tilde{E}_v\rangle$ is $|g\rangle$-type state, Eq.(5). This is also the case for the ADS [8].

The virtual components of the NADS (and these of ADS) are *real physical states*, but not simply a mathematical construct, due to the following arguments: (*i*) real population on the virtual components has been observed experimentally like in any other real quantum state and it can be transferred to the real component [6]; (*ii*) the real components "feel" the appearance of the virtual components in their vicinity as the appearance of any other real state. Thus, if the excitation due to the external field starts from $|\tilde{G}_r\rangle$ and the created virtual state $|\tilde{G}_v\rangle$ is at the low energy side of state $|\tilde{E}_r\rangle$, Fig.1, the real state $|\tilde{E}_r\rangle$ is "pushed" toward the high energy side. At the same time, for the excitation starting from $|\tilde{E}_r\rangle$, the position of the created virtual state $|\tilde{E}_v\rangle$ is at the high energy side of state $|\tilde{G}_r\rangle$, Fig. 1, and the real state $|\tilde{G}_r\rangle$ is "pushed" toward the low energy side. Opposite shift takes place if the created virtual ground state $|\tilde{G}_v\rangle$ is at the high energy side of $|\tilde{E}_r\rangle$, while the created virtual excited state $|\tilde{E}_v\rangle$ is at the low energy side of $|\tilde{G}_r\rangle$. Hence, the virtual state and nearest real state, *e.g.*, $|\tilde{G}_v\rangle$ and $|\tilde{E}_r\rangle$, as well as $|\tilde{E}_v\rangle$ and $|\tilde{G}_r\rangle$, repeal each other. Note that $|\tilde{G}_v\rangle$ and $|\tilde{E}_r\rangle$ are both $|e\rangle$-type states and $|\tilde{E}_v\rangle$ and $|\tilde{G}_r\rangle$ are both $|g\rangle$-type states. In the other words, the states of same symmetries repeal each other. The repulsion of states of same symmetry is well known for quantum systems with internal degrees of freedom (*e.g.*, states of diatomic molecules with internuclear distance as internal degree of freedom) in accordance with von Neumann-Wigner non-crossing rule [10]. The repulsion of NADS components, and those of ADS, can be considered as a manifestation of the non-crossing rule for quantum systems without internal degrees of freedom and shows that the virtual components appear as real physical states.

The BS, from which the real components of the NADS originate, are stationary states of completely isolated (closed) quantum system and their lifetime should be infinity. In reality, due to various perturbations, but at least, the zero point vacuum fluctuations, only the ground state has infinite lifetime, while all excited states have a limited lifetime. *The BS are only the stable states in a closed quantum system*. Due to same reason, *the real components of the NADS*, as they originate from the respective BS, *are only the stable states in an open quantum system*. The lifetime of real NADS may differ from that of BS due to the external perturbations. The virtual components of the NADS are unstable and exist only during the supporting action of the external electromagnetic field. At zero field, the virtual components of NADS, and those of ADS, disappear. Consequently, the population, *e.g.*, the electron, may reside on the real NADS within the lifetime of the state even after the external field is switched off, whereas the population, *e.g.*, the electron, may reside on the virtual NADS only during the existence of the external field.

Some quantum processes are characterized as "simultaneous", "instantaneous", "jumps", etc., but these have never been confirmed experimentally with the required time resolution, especially what concerns light particles as electrons. The typical characteristic time of electron motion in atoms falls within the attosecond time scale, $1as = 10^{-18} s$. The recent attosecond technology is close to this time domain but, to the best of our knowledge, the exact timing of superposition of the quantum states has not been yet studied experimentally. That is why, our further considerations will be based on a remarkable experimental work, which, while having only nanosecond time resolution, reveals the dynamics of population of the quantum states during the quantum transitions [4]. To make the relation with this work closer, we will apply our notations to the energy level scheme involved in these experiments, Fig.2. The atom is excited adiabatically at frequency ω from the real ground NADS $|\tilde{G}_r\rangle$ (or $|G_r\rangle$ for ADS) to the virtual ground state $|\tilde{G}_v\rangle$ ($|G_v\rangle$), which is $\Delta\omega = 0.8 cm^{-1}$ off resonance with the real excited state $|\tilde{E}_r\rangle$ ($|E_r\rangle$). At the same time, the Doppler width of $|\tilde{E}_r\rangle$ state and the spectral width of the exciting laser pulse are $\delta\omega_D = 0.04 cm^{-1}$ and $\delta\omega_L = 0.005 cm^{-1}$, respectively, or adiabatic condition is well satisfied in its frequency form, $\Delta\omega \gg \delta\omega_D, \delta\omega_L$. Thus, no spectral components of the exciting laser cover $|\tilde{G}_r\rangle \to |\tilde{E}_r\rangle$ transition and only population of $|\tilde{G}_v\rangle$, but not of $|\tilde{E}_r\rangle$, must be expected according to the adiabatic condition. However, this is adiabatic condition only for the field. The experiment is done at $3.10^{13} cm^{-3}$ atomic density and more than $400K$ temperature. This means that, while all nonadiabatic factors are excluded in theory from the ADS, the nonadiabatic factors from the environment (mainly collisions with other atoms) cannot be excluded in the experiment and they are mainly responsible for the population of the real excited state $|\tilde{E}_r\rangle$. Such experimental conditions have crucial importance so as to distinguish adiabatic population of the virtual state $|\tilde{G}_v\rangle$ from the nonadiabatic population of real state $|\tilde{E}_r\rangle$. The population on the real state $|\tilde{E}_r\rangle$ is measured by a probe light of frequency $\omega_{p1}$, which is resonant with $|\tilde{E}_r\rangle \to |n\rangle$ transition, while the

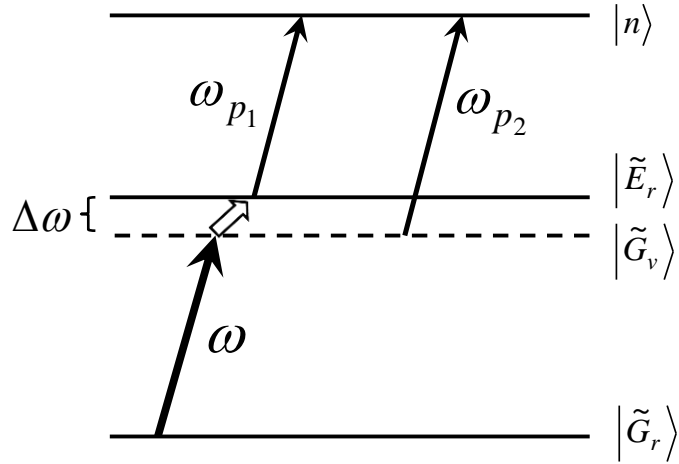

**FIGURE 2.** Energy level diagram used in Ref. 6 for measurement of incoherent and coherent population of $|\tilde{E}_r\rangle$ and $|\tilde{G}_v\rangle$, respectively, but depicted in the notations used in the present work.

population on the virtual state $|\tilde{G}_v\rangle$ is measured by a probe light of frequency $\omega_{p2}$, which is resonant with $|\tilde{G}_v\rangle \to |n\rangle$ transition, where $|n\rangle$ is a higher lying state, Fig.2. The experiment shows that the peak population of the virtual ground state $|\tilde{G}_v\rangle$ is more than an order of magnitude higher than that of real state $|\tilde{E}_r\rangle$. Also, the population of $|\tilde{G}_v\rangle$ is *coherent*, its instantaneous value is proportional to the time shape of intensity of the pump pulse, while the population of $|\tilde{E}_r\rangle$ is *incoherent* and proportional to the time integral of intensity of the pump pulse [4]. All these, together with the fact that exciting laser is well off resonant with $|\tilde{E}_r\rangle$ state, lead to conclusion that first $|\tilde{G}_v\rangle$ is populated and after that the population is transferred to $|\tilde{E}_r\rangle$ state. Hence, there are *two population mechanisms*, which are *separated in time*. They will be summarized for the processes of absorption and emission, Fig.1:

*absorption*

*first step*: formation of a virtual ground state from the real ground state due to *coherent* excitation from the field, $|\tilde{G}_r\rangle \rightarrow |\tilde{G}_v\rangle$. This process develops entirely within the ground NADS $|\tilde{G}\rangle$, $|\tilde{G}\rangle = COS(\theta/2)|\tilde{G}_r\rangle + SIN(\theta/2)|\tilde{G}_v\rangle$.

*second step*: *nonadiabatic incoherent* transition from the virtual ground state to the real excited state $|\tilde{G}_v\rangle \Rightarrow |\tilde{E}_r\rangle$. The process is accompanied by irreversible absorption of one photon from the field and terminates with population of $|\tilde{E}_r\rangle$. In essence, the process represents transition between two different NADS, ground $|\tilde{G}\rangle$ and excited $|\tilde{E}\rangle$.

*emission (stimulated)*

*first step*: formation of a virtual excited state from the real excited state due to *coherent* excitation from the field, $|\tilde{E}_r\rangle \rightarrow |\tilde{E}_v\rangle$. This process develops entirely within the excited NADS $|\tilde{E}\rangle$, $|\tilde{E}\rangle = COS(\theta/2)|\tilde{E}_r\rangle - SIN(\theta/2)|\tilde{E}_v\rangle$.

*second step*: *nonadiabatic incoherent* transition from the virtual excited state to the real ground state $|\tilde{E}_v\rangle \Rightarrow |\tilde{G}_r\rangle$. The process is accompanied by irreversible emission of one photon to the field and terminates with population of $|\tilde{G}_r\rangle$. In essence, the process represents transition between two different NADS, excited $|\tilde{E}\rangle$ and ground $|\tilde{G}\rangle$.

If, as in most cases of resonant excitation, the light is resonant $\Delta\omega = 0$ with given transition $|\tilde{G}_r\rangle \rightarrow |\tilde{E}_r\rangle$, the adiabatic condition is violated, and these two steps cannot be distinguished - the transition appears as a single step process.

One absorption and one emission process between two real components of NADS are separated in time because, as explained above, they are adiabatically decoupled. They may form a closed cycle, *nonadiabatic loop*, $|\tilde{G}_r\rangle \rightarrow |\tilde{G}_v\rangle \Rightarrow |\tilde{E}_r\rangle \rightarrow |\tilde{E}_v\rangle \Rightarrow |\tilde{G}_r\rangle$, Fig.1, or even several loops, within given excitation of the field. In this sequence, full arrow stands for the usual radiative process and hollow arrow stands for the nonadiabatic process. The time duration of the nonadiabatic loop, $\tau_{NAL}$, is non-zero and depends mainly on the intensiveness of the nonadiabatic factors that transfer the population between different NADS, because the electronic transition in the radiative part of the loop has, probably, attosecond time duration.

## 3.2 Quantum Superposition

One of the basic principles in quantum mechanics is the *quantum superposition principle*, according to which, if a number of states are physical states of given quantum system, *i.e.*, they obey the Schrödinger equation, any linear superposition of these states is also a state of that quantum system. It, of course, is correct from a formal mathematical point of view, because of the linearity of the Schrödinger equation. Such a superposition of states is called *coherent* and allows observing quantum interference effects. As will be seen below, the physical properties of NADS have crucial importance to understand the superposition process in quantum systems, and, both together, will help to understand the collapse of wave function and quantum measurement problem.

For quantum systems as atoms, molecules, etc., it is usually considered (while not always) that the states that are superimposed are the usual BS. In the two-level system, the superposition state is:

$$|\psi(\vec{r},t)\rangle = c_1(t)|g(\vec{r},t)\rangle + c_2(t)|e(\vec{r},t)\rangle \qquad (7)$$

where it is automatically accepted that the BS $|g\rangle$ and $|e\rangle$ are superimposed *simultaneously*. The ground and excited NADS (but it also holds for the ground and excited ADS) are *adiabatically decoupled* and the quantum system evolves only within given NADS, if the nonadiabatic factors from the field and the environment can be neglected. Transitions between different NADS originate from the nonadiabatic factors acting on the quantum system. This, of course, is in agreement with *adiabatic theorem of quantum mechanics* [11], according to which, if the Hamiltonian of the quantum system changes slowly enough, *i.e.*, adiabatically, the quantum system has time to rearrange and it passes from the initial eigenstate of the nonperturbed Hamiltonian through a continuous succession of eigenstates of the perturbed instantaneous Hamiltonian and, once the perturbation terminates, the quantum system becomes again in the same initial eigenstate of the nonperturbed Hamiltonian, and no transition to other states will occur. Based on above theoretical and experimental results, it is quite natural to raise the question: "How superposition from different stationary states, *i.e.*, BS (which are usually considered in the quantum superposition, Eq. (7)) or their perturbed counterpart - the real components of NADS or ADS, can be *simultaneous* and *coherent* if such states are adiabatically decoupled and transition between these states results from a secondary well distinguished in time incoherent nonadiabatic process?". Due to above evidences, we have to accept that superposition of two BS, $|g\rangle$ and $|e\rangle$, as in Eq. (7), or their perturbed counterpart - two real components of NADS $|\tilde{G}_r\rangle$ and $|\tilde{E}_r\rangle$, or ADS $|G_r\rangle$ and $|E_r\rangle$, cannot be simultaneous and coherent (see the preceding section where the steps of emission and absorption processes are characterized). *Coherent superposition* will consist of given real state and the corresponding virtual state, *e.g.*, $|\tilde{G}_r\rangle$ and $|\tilde{G}_v\rangle$ within ground NADS, or $|\tilde{E}_r\rangle$ and $|\tilde{E}_v\rangle$ within excited NADS (the same holds for ADS):

$$|\tilde{G}\rangle = COS(\theta/2)|\tilde{G}_r\rangle + SIN(\theta/2)|\tilde{G}_v\rangle$$
$$|\tilde{E}\rangle = COS(\theta/2)|\tilde{E}_r\rangle - SIN(\theta/2)|\tilde{E}_v\rangle \quad (8)$$

Note that the superposition in $|\tilde{G}\rangle$ also consists of one $|g\rangle$-type state ($|\tilde{G}_r\rangle$) and one $|e\rangle$-type state ($|\tilde{G}_v\rangle$), Eq.8, but, in contrast to Eq. (7), the $|e\rangle$-type state in Eq. (8) is virtual and coherent with $|g\rangle$-type state. The same holds for the superposition in $|\tilde{E}\rangle$, which also consists of, this time, one $|e\rangle$-type state ($|\tilde{E}_r\rangle$) and one $|g\rangle$-type state ($|\tilde{E}_v\rangle$), where $|g\rangle$-type state in (8) is virtual and coherent with $|e\rangle$-type state. For completeness, we cannot exclude the existence of non-instantaneous, *i.e.*, sequential, incoherent superposition of BS, or real components of different NADS or ADS, within the time duration of given nonadiabatic loop $\tau_{NAL}$. The second part of the question still remains open: will such coherent superposition within given NADS be simultaneous or sequential, *i.e.*, will electron resides on the real and virtual component within given NADS simultaneously, or it goes around real and virtual components sequentially. The simultaneous occupation of few quantum states is well accepted in quantum mechanics, but the second possibility cannot be excluded until the problem is solved by a proper experiment. As can be seen from the present considerations, the coherent superposition of quantum states (at least for quantum systems under considerations - atoms, molecules, etc.) is a causal process, whose mechanism can be well traced.

Finally, we have to say that the ordinary superposition, Eq. (7), has been used as a starting point in the derivation of NADS [5] because the correct final result cannot be presupposed at a not yet solved problem. This, however, does not change our conclusions because they are supported also by a number of theoretical and experimental arguments.

### 3.3 Collapse of Wave Function and Quantum Measurement Problem

In the preceding section we have considered how to create coherent superposition state, based on the physical properties of NADS. The collapse of the wave function and the quantum measurement represent opposite process – *destruction of coherent superposition* in a way that the quantum system to be in *only one* of its eigenstates. Here, we consider the physical mechanism to explain the collapse of wave function and quantum measurement taking into account the mechanism of quantum superposition, proposed above, and the physical properties of NADS.

In agreement with above conclusions, the quantum system may go around *sequentially* between two different NADS (ground and excited, for the two-level case) within given nonadiabatic loop(s) of nonzero time duration, but it may occupy only one of these NADS at given instant of time because the different NADS are adiabatically decoupled. Also, as transition between different NADS results from the action of incoherent nonadiabatic factors, sequentially populated different NADS are not coherent. This rule is automatically transferred to the real components of the different NADS (ADS), the quantum system cannot be simultaneously and coherently in two real component of different NADS (ADS). Finally, at zero field limit and closed quantum system, the NADS and the ADS evolve toward the BS, and the quantum system cannot be simultaneously and coherently in different BS, *i.e.*, the different BS are also adiabatically decoupled. On the other hand, the quantum system, being in given NADS, is in coherent superposition of the real and the virtual components of this NADS, subject to coherence of the creating field - see the exponential phase factors of the NADS components in Eq.5. The BS, or their perturbed counterpart - the real components of NADS, or the AND, are only the stable states in a given quantum system. The quantum system may reside in these states within their lifetime even after the external field is switched off. When the field is switched off, the virtual component disappears, while the real component of the NADS evolves toward the respective BS, and the NADS ceases to exist. Thus, only the BS (ground and excited, for the two-level case), from which the respective NADS (ground or excited) originate, continues existing and the quantum system can be trapped in some of the BS within the lifetime of this state. This leads us to the following understanding about the collapse: *the collapse of the wave function results from trapping the quantum system in only one of the BS once the external field is switched off*. Thus, the collapse of the wave function is a real physical process with superimposed quantum states where the external perturbations simply help this to occur.

In agreement with the above discussion, the ordinary scheme of collapse of superposition state composed of BS, *i.e.*, $|\psi\rangle = c_1|g\rangle + c_2|e\rangle \to |g\rangle$ or $|\psi\rangle = c_1|g\rangle + c_2|e\rangle \to |e\rangle$, which is not simultaneous and coherent, will be replaced by collapse of NADS, which is a simultaneous and coherent superposition of real and virtual components:

$$\begin{aligned} |\tilde{G}\rangle &= COS(\theta/2)|\tilde{G}_r\rangle + SIN(\theta/2)|\tilde{G}_v\rangle \to |g\rangle \\ |\tilde{E}\rangle &= COS(\theta/2)|\tilde{E}_r\rangle - SIN(\theta/2)|\tilde{E}_v\rangle \to |e\rangle \end{aligned} \qquad (9)$$

In the case of quantum measurement, a quantum system *S* (usually microscopic) interacts by means of a given field with a measurement apparatus *A* (usually macroscopic but considered quantum mechanically) and the surrounding environment *E*. This physics is naturally included, in principle, in the NADS by means of the respective three parts of the Hamiltonian, Eq.(1). The quantum measurement problem consists in the fact that, each time measurement on a given observable takes place, different but well defined value of the measured quantity is obtained - the eigenvalue of only one of the eigenstates of the quantum system on which the wave function collapses, even if arbitrary number of eigenstates of the quantum system is involved. To explain the quantum measurement, we will apply the above scheme of collapse to the measurement process. When two quantum systems, S and A, of eigenstates (for the two-level case) denoted by $|g\rangle$, $|e\rangle$ and $|A_g\rangle$, $|A_e\rangle$, respectively, interact, they may form entangled state of the combined system $|\psi_{SA}\rangle$, which has the following structure $|\psi_{SA}\rangle = |g\rangle|A_g\rangle + |e\rangle|A_e\rangle$. If a measurement takes place, the entangled state must be destroyed and the quantum system will be found either in $|g\rangle$ or in $|e\rangle$, and this exactly correlates with the state in which the measuring apparatus will be found - in $|A_g\rangle$ or in $|A_e\rangle$, respectively. However, as we said above, if $|g\rangle$ and $|e\rangle$ are BS, the quantum system cannot be simultaneously and coherently in both BS and such entangled state cannot exist at a given instant of time. Therefore, as for the case of collapse, the ordinary scheme of entangled state composed from BS must be replaced by scheme based on NADS, ground NADS Eq.(10) or excited NADS Eq.(11):

$$|\psi_{SA}\rangle = |\psi_{\tilde{G}A}\rangle = COS(\theta/2)|\tilde{G}_r\rangle|A_g\rangle + SIN(\theta/2)|\tilde{G}_v\rangle|A_e\rangle \to |g\rangle|A_g\rangle \qquad (10)$$

$$|\psi_{SA}\rangle = |\psi_{\tilde{E}A}\rangle = COS(\theta/2)|\tilde{E}_r\rangle|A_e\rangle - SIN(\theta/2)|\tilde{E}_v\rangle|A_g\rangle \to |e\rangle|A_e\rangle \qquad (11)$$

The process of measurement represents an interaction between the quantum system and the measuring apparatus by means of a given field, in this case, electromagnetic field. The electromagnetic field of the measuring apparatus is the field that now creates NADS from the initial BS. The NADS are in fact entangled states between the states of the quantum system and the field of the measuring apparatus. This is more clearly seen for the case of quantized fields. As the states of the quantum system and the apparatus field are combined and entangled within the NADS, the states $|A_g\rangle$ and $|A_e\rangle$ here play role of pointer states of the apparatus that follow the evolution of the states of the field.

The quantum measurement can now be easily explained. Within given nonadiabatic loop(s), the quantum system may go around sequentially between different NADS, ground and excited. If the field of interaction between the quantum system and the measuring apparatus is switched off, or the quantum system leaves the area of the measurement field (different realistic measurement schemes can also be proposed), *i.e.*, the measurement terminates, the quantum system will be either in $|\tilde{G}\rangle$ or in $|\tilde{E}\rangle$, because it cannot be simultaneous in both NADS. Thus, only one of the entangled states (10) or (11) takes place at given instant of time. If the field becomes zero, both virtual components of the NADS disappear, $|\tilde{G}_v\rangle \to 0$, $|\tilde{E}_v\rangle \to 0$, whereas the real components of the NADS evolve toward the respective BS, $|\tilde{G}_r\rangle \to |g\rangle$, $|\tilde{E}_r\rangle \to |e\rangle$. Thus, the quantum system will be found in only one of the BS because the different BS originate from different NADS, in which the quantum system cannot be simultaneously but only sequentially. According to Eqs.(10) and (11), only one of the states, $|A_g\rangle$ or $|A_e\rangle$, of the measuring apparatus will survive, thus showing which state the quantum system is in at the end of the measurement, $|g\rangle$ or $|e\rangle$, respectively.

## 3.4 Quantum Jumps

The dressed state picture is very useful to explain the problem with the so called *"quantum jumps"*. The idea of quantum jumps arises from the quantization of atomic energy levels, which dates since the old Bohr's atomic model. According to it, only well-defined atomic levels of quantized energies are allowed, while the intermediate energies are forbidden. Thus, if electron makes a transition from one level to the other level, it must be in the form of abrupt transition - quantum jump. As has been already mentioned, the characteristic time of electron motion is in the attosecond time domain and no real experiment has been done that can trace directly such a motion. Some experiments with single atoms [12, 13] claim to observe quantum jumps, which are expressed in the fact that observed fluorescence abruptly disappears and appears again. However, this behavior of the fluorescence is displayed in a time scale of seconds and the detection methods used do not allow detecting electron processes in real time in atoms. Recently, it has been shown [14] that the evolution of the quantum transitions is continuous, coherent and deterministic, and it can be controlled and even reversed. Although such a possibility is highly intriguing, the experiment is not done with real atoms but the so called "artificial atom" (a qubit fabricated by microelectronic technology) is used. On the other hand, the experiments with real atoms [6, 7], although not done with time resolution to resolve the electron motion, show that the quantum transition, *e.g.*, the usual single photon absorption between two BS, is not an instantaneous process, *i.e.*, not a quantum jump, but represents a well traceable two-step process [5]. The first step consists of formation of a virtual component of the dressed state under the action of external electromagnetic field. The virtual state can be formed everywhere on the energy scale and its position depends on the photon energy of the field. The electron resides on the virtual state for a given nonzero time, which is the reason to detect a real population on the virtual state [6, 7]. The second step consists in the nonadiabatic transition from the created virtual state to the final real state under the action of nonadiabatic factors. All these not only show that the quantum transition between two quantum states is not a jump but, in addition, reveal the complicated internal dynamics of the quantum transitions.

## 4. CONCLUSION

Quantum superposition, collapse of wave function and quantum measurement problem are considered within the nonadiabatic dressed state picture and experimental investigations on the quantum transitions within two-level quantum systems. The coherent quantum superposition is explained as a superposition of real and virtual components within given NADS. The collapse of the wave function and the quantum measurement problem are explained as trapping of the quantum system in a given real component of the nonadiabatic dressed state once the external perturbation, *e.g.*, electromagnetic field, is switched off. Switching the field off breaks the superposition, *i.e.*, the dressed state, and the latter collapses on the respective real component. Within the dressed state picture, the quantum transition between two real quantum states does not have behavior of "quantum jump" but represents a well traceable two-step process, neither of which is instantaneous.